\documentclass[12pt]{iopart}
\usepackage{bm,amsfonts,amssymb, graphicx}
\usepackage{epstopdf}

\begin{document}

\title{Luminescence properties of the hybrid Si-Ni nanoparticles system}
\author{A. A. Lalayan, S. S. Israelyan, H. A. Movsesyan}

\address{Centre of Strong Fields Physics, Yerevan State University, 1 A. Manoogian,
Yerevan 0025, Armenia}
\ead{alalayan@ysu.am}

\begin{abstract}
The luminescence properties of the colloidal hybrid Si -- Ni nanoparticles
system fabricated in the pure water by pulsed laser ablation is considered.
The red-shifted photoluminescence of this system because of the Stark effect
in the Coulomb field of the charged Ni nanoparticles has been registered in
the blue range of the spectrum.
\end{abstract}

\pacs{73.21.La, 42.62.Fi, 61.46.Hk}


\vspace{2pc} \noindent\textit{Keywords}: Laser ablation, colloidal quantum
dots, Si-Ni nanoparticles, photoluminescense, Stark shift 
\maketitle
\section{Introduction}

Nanoparticles (NPs) with sizes from a few nanometers to tens of nanometers
are artificial materials which exhibit extraordinary optical and electrical
properties due to its ultra-small sizes and in particular to quantum
confined nature of their energy levels.

The extraordinary properties of size-tunable NPs has given rise to their
wide spread applications in nanophotonics, plasmonics, biomedicine, lasing,
photovoltaic etc \cite{1,2,3}. Semiconductor NPs or quantum dots (QDs) are
found a number of applications in the biomedical imaging due to the size-
dependent flexibility of its luminescent properties. The NPs with the sizes
2-10nm, which is about of the size of small biomolecules, have the great
potential for various applications in the modern biomedicine. In difference
from organic biological labels, the luminescent QDs have several advantages.
At first they have a wide absorption band and can be used with a variety of
laser sources having different output wavelengths. On the other side, the
luminescence band of QDs can be tuned from UV to IR by changing their sizes
and this property allows combining several types of QDs at the one
excitation wavelength. It is also important to notice that QDs are more
photostable under the long time of laser irradiation. The existence of the
mentioned properties of QDs could dramatically improve the use of
fluorescent markers in the biological imaging.

During the last decade various methods have been investigated for the
synthesis of NPs. Recently, a simple and flexible method of the pulsed laser
ablation in liquid media (PLAL) has been demonstrated for the laser
synthesis of metal nanoparticles \cite{4,5,6}. This method also has been
applied for synthesis of colloidal solutions of semiconductor QDs \cite%
{7,8,9}. Laser ablation in liquids media is especially good-looking for
biomedical applications, since NPs could be formed directly in specially
prepared bio-conjugated media and such NPs colloidal solution can be exactly
tuned for the specific biological target.

The formation of semiconductor nanoparticles via PLAL has been investigated
in different experimental conditions, and photoluminescence properties of
nanoparticles produced by the PLAL have been found to be strongly dependent
as on the physical and chemical properties of liquid environment, as well as
on the parameters of the laser irradiation. Laser ablation syntheses of
nanoparticles could be explained in terms of the dynamic formation mechanism
of the particle growth \cite{10}. According this mechanism, a material of
target evaporated in the laser plume very quickly aggregate into small
embryonic particles, therefore such laser pulse parameters as pulse energy
and duration could be highly critical in this process. In the paper \cite{11}
the influence of a laser beam transverse electromagnetic mode structure on
the luminescence properties of laser synthesized GaAs nanoparticles has been
studied. In the present work we studied luminescence properties of the Si --
Ni hybrid colloidal nanoparticles system fabricated in pure water by the
laser ablation method using a nanosecond Nd:YAG laser.

\section{Experimental arrangements}

The pulsed passive Q-switched Nd:YAG laser with wavelength of 1064nm, pulse
duration of 8ns and repetition rate of 10Hz was used to synthesize
nanoparticles in pure water. The laser beam was focused on the surface of a
bulk target (the plate of bulk Si or Ni) allocated in the glass cuvette with
distilled water (see Figure (1)). Exposition of the laser irradiation was
two hours. To produce hybrid metal-semiconductor system, at first we
irradiated a Ni bulk target, then in the same cuvette with water solution of
Ni nanoparticles was located the Si bulk target to produce Si nanoparticles.
Hybrid nanoparticles system was produced also in similar scheme when Ni and
Si nanoparticles were produced separately in different cuvettes and after
that mixed in one solution. Note that luminescence spectra of the obtained
hybrid nanoparticles were the same in these two experimental approaches.
Note also that a luminescence of separately prepared Si nanoparticles was
not observed. 
\begin{figure}[tbp]
\begin{center}
\includegraphics{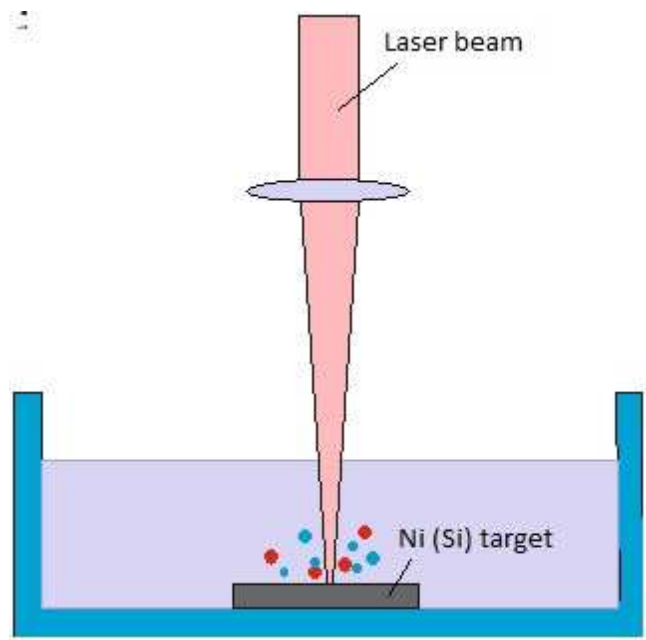} \label{fig1}
\end{center}
\caption{Schematic presentation of the laser synthesis of colloidal
nanoparticles in the liquid medium.}
\end{figure}
For luminescence measurements the water solution of nanoparticles was
relocated into quartz cuvette with the transparent walls in the UV spectral
range. The luminescence of the colloidal QDs was excited by the irradiation
at 400 nm wavelength of the continuous wave diode laser with the power of
10mW. The luminescence spectra were analyzed by spectrometer constructed on
the base of monochromator LOMO MDR- 23-PMT. Electrical signal amplifier was
used to enhance output of the PMT. Luminescence spectrum of colloidal Si -
Ni QDs is presented in the Figure (2). 
\begin{figure}[tbp]
\begin{center}
\includegraphics{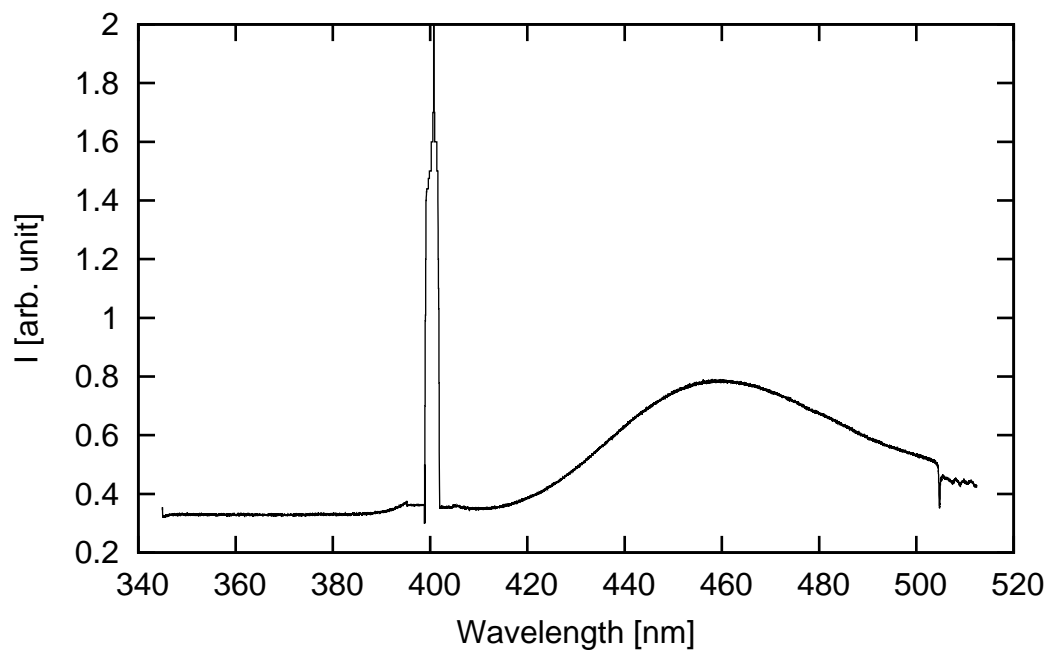} \label{fig2}
\end{center}
\caption{Photoluminescence spectrum of colloidal Si - Ni nanoparticles.}
\end{figure}
The considerable blue shift of the photoluminescence has been registered
with the location of the maximum at the wavelength of 455nm which is
connected with the size-effect and significant difference from the well
known luminescence of a bulk Si material located in the near infrared
region. The peak on the 400nm is the excitation laser wavelength. The
bandwidth of the luminescence spectra on the halfheight was about 40nm that
demonstrates the narrow size distribution of laser generated QDs, which is
achieved without application of any size selection schemes. 
\begin{figure}[tbp]
\begin{center}
\includegraphics{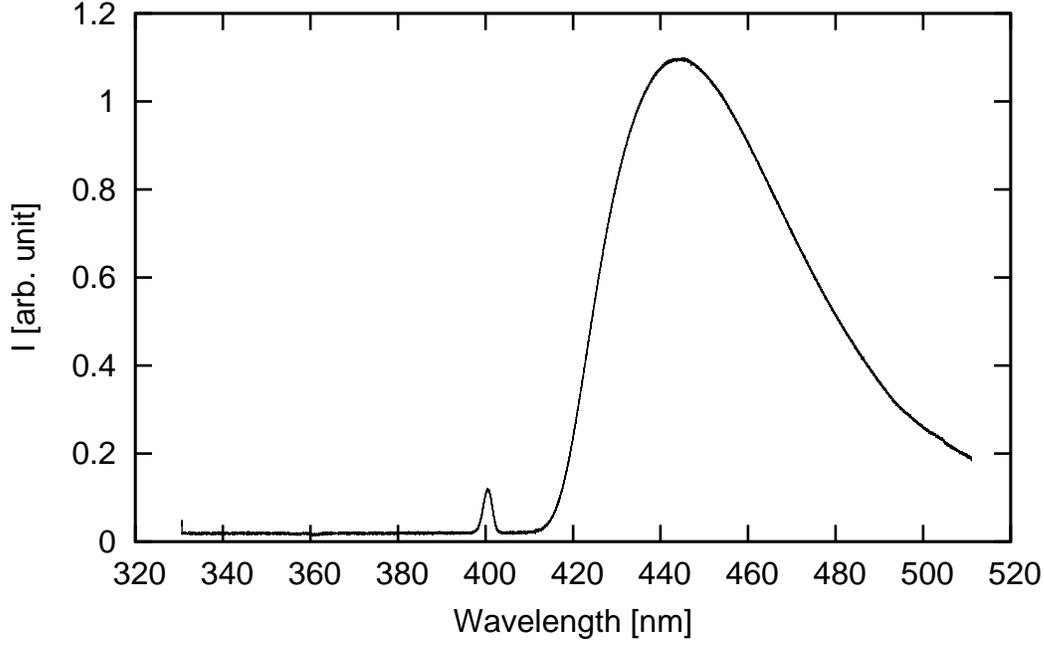} \label{fig3}
\end{center}
\caption{Photoluminescence spectrum of the Coumarin 120 organic dye (water
solution).}
\end{figure}
Possible contributions to the luminescence from a quartz material of the
cuvette and water has been eliminated by a comparison with the
photoluminescence from the same cuvette filled with the clean water or with
the solution of Coumarin 120 organic dye (see Figure (3)). The long-term
stability of the luminescence properties of the produced QDs was observed
during the one year.

\section{Theoretical consideration and discussion}

Here we give some interpretation of the results of our experiment by
estimations based on the known properties of silicon nanoparticles and
quantum confinement effect, as well as the effect of the charged nickel
nanoparticles on luminescence spectra of the colloidal Si quantum dots.

As is known \cite{12}, metalic nanoparticles generated in diverse media by
the method of a laser ablation carry a positive or negative electrical
charge. In our case, the charged Ni nanoparticles within the Si -- Ni hybrid
system induce a static electrical field which causes Stark shift of the
electron-hole energy levels in the silicon QD. In the following, we consider
a point charge at a certain distance from the surface of the silicon QD (see
Figure (4)). The assessment of the Si nanoparticles luminescence spectrum
shifting under the influence of a static electric field of an external
charge allows us to explain experimental results by the Stark effect in the
inhomogeneous (Coulomb) field.

The external point charge $q$ on the distance $R_{q}$\ from the center of
the silicon quantum dot creates a screened electrostatic field at the point $%
\mathbf{r}$ inside the QD, potential of which can be given in the following
form \cite{13}:

\begin{equation}
\Phi \left( \mathbf{r}\right) =q\sum\limits_{n=0}^{\infty }\frac{2n+1}{%
\epsilon n+n+1}\frac{r^{n}}{R_{q}^{n+1}}P_{n}\left( \cos \vartheta \right) ,
\label{1}
\end{equation}%
where $P_{n}\left( \cos \vartheta \right) $ is the Legendre Polynomial, $%
\epsilon $ -screening dielectric constant of Si QDs, $\vartheta $ -angle
between the vectors $\mathbf{r}$ and $\mathbf{R}_{q}$. The shift of the
energy levels of an exciton caused by an external electrostatic field can be
evaluated using the wavefunction $\Psi _{0}\left( \mathbf{r}_{e},\mathbf{r}%
_{h}\right) $ of the electron-hole system:

\begin{equation}
\bigtriangleup E=\int d\mathbf{r}_{e}d\mathbf{r}_{h}\left[ \Phi\left( 
\mathbf{r}_{h}\right) -\Phi\left( \mathbf{r}_{e}\right) \right] \left\vert
\Psi_{0}\left( \mathbf{r}_{e},\mathbf{r}_{h}\right) \right\vert ^{2}.
\label{2}
\end{equation}

Taking into account the form of the potential of the external field (\ref{1}%
), the energy shift (\ref{2}) can be represented in the form of the sum

\begin{equation}
\bigtriangleup E=\sum \limits_{n=0}^{\infty}\bigtriangleup E_{n},  \label{3}
\end{equation}
where

\[
\bigtriangleup E_{n}=\frac{eq}{R_{q}}\int d\mathbf{r}_{e}d\mathbf{r}%
_{h}\left\{ \frac{2n+1}{\epsilon n+n+1}\right. 
\]%
\begin{equation}
\times \left. \frac{\left( r_{h}^{n}P_{n}\left( \cos \vartheta _{h}\right)
-r_{e}^{n}P_{n}\left( \cos \vartheta _{e}\right) \right) }{R_{q}^{n}}%
\left\vert \Psi _{0}\left( \mathbf{r}_{e},\mathbf{r}_{h}\right) \right\vert
^{2}\right\} .  \label{4}
\end{equation}

The first term in the sum (\ref{3}): $\bigtriangleup E_{0}=0$. We will
estimate only the term corresponding to $n=1$ taking into account the
smallness of the other terms. Thus,

\begin{equation}
\bigtriangleup E_{1}=\frac{eq}{R_{q}}\int d\mathbf{r}_{e}d\mathbf{r}_{h}%
\frac{3}{\epsilon +2}\frac{r_{h}\cos \vartheta _{h}-r_{e}\cos \vartheta _{e}%
}{R_{q}}\left\vert \Psi _{0}\left( \mathbf{r}_{e},\mathbf{r}_{h}\right)
\right\vert ^{2},  \label{5}
\end{equation}%
which can be written in the following form:

\begin{equation}
\bigtriangleup E_{1}=\frac{Deq}{R_{q}^{2}}\frac{3}{\epsilon+2}\left( \frac{%
\left\langle z_{h}\right\rangle -\left\langle z_{e}\right\rangle }{D}\right)
,  \label{6}
\end{equation}
where $D$ is the size of an exciton, $\left\langle z_{h,e}\right\rangle
=\int\left\vert \Psi_{0}\left( \mathbf{r}_{e},\mathbf{r}_{h}\right)
\right\vert ^{2}z_{h,e}d\mathbf{r}_{e}d\mathbf{r}_{h}$ are the mean values
of the charge distribution density of the hole and the electron,
respectively. To calculate the energy spectral shifting in accordance with
the conditions of the above described experiment, let us estimate $%
\left\vert \bigtriangleup E_{1}\right\vert $.

As is shown in \cite{14}, the emitted photon energy depends on silicone
nanocrystallite size. With the decreasing of the size of nanoparticles the
quantum confinement affects on the band gap of the exciton thereby
increasing emission energy. During the experiment with the solely silicon
nanoparticles the luminescence had not been observed in the visible spectrum
at the irradiation by the wavelength of 400 nm. This fact evidences that
bandgap energy of the Si QDs is more than 3 eV, which is agreed with the
results of the work \cite{15}. According to the work \cite{14}, such
energies corresponds to silicon nanoparticles with diameter less than 1.4
nm. To determine the other parameters we will take for screening dielectric
constant $\epsilon =6$, which corresponds to the diameter $D=1.3$ nm of Si
quantum dots at the passivation with the hydrogen atoms, according to the
paper \cite{16}.

As we consider Si nanoparticles generated in a liquid medium (pure water in
our case), dangling bonds of the silicon atoms on the dot surface should be
saturated mostly by the hydrogen atoms. Note that in the process of the
laser ablation hydrogen molecules are produced due to water molecules
dissociation in the surroundings of the laser plume. As is shown in the work 
\cite{16}, in the case of the hydrogen passivated Si nanoparticles the size
of the exciton is approximately equal to the diameter of the QD.

\begin{figure}[tbp]
\begin{center}
\includegraphics{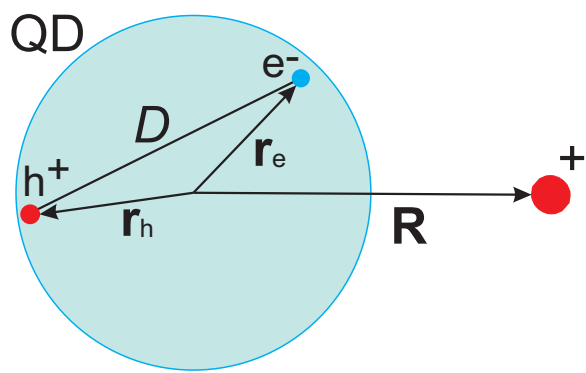} \label{fig4}
\end{center}
\caption{Schematic presentation of the point charge interaction with the
exciton at the distance R from the QD center.}
\end{figure}

Since we consider the nickel nanoparticles in close proximity from the
silicone QDs, then $R_{q}-R\ll R$, and Coulomb potential is strong enough to
pull the hole away from the electron. In this case, the quantity $\left\vert
\left\langle z_{h}\right\rangle -\left\langle z_{e}\right\rangle \right\vert
/D$ can reach as large as $0.5$ \cite{17}. Then, we have $\left\vert
\bigtriangleup E_{1}\right\vert =0.207$ eV. This shift of bandgap energy
corresponds to a red shift of Si quantum dot luminescence spectrum by about $%
30$ nm (it is clear that depending on the distance of the point charge from
the surface of silicon QD this shift will be changed). This shift can
explain the observation of blue luminescense in the case of hybrid Si-Ni
nanoparticles system in opposite to the case of the solely Si nanoparticles
in the colloidal solution.

Summarizing, by the luminescense picture of the mixed metallic-semiconductor
nanoparticles in the colloidal solution one can define the sizes of quantum
dots. The Si colloidal QDs in liquids can be used in a number of optical
applications, in particular, as luminescent markers in the biomedical
imaging in vivo and in vitro. The knowledge of peculiarities of their
luminescent properties is very important for such kinds of biomedical
applications. More complex assemblies of nanocrystals can be easily
constructed by the presented technique of laser synthesis of colloidal QDs
including complexes of nanoparticles of different materials. These complex
artificial structures will exhibit new physical and chemical properties that
require further investigation and could be applied in the optical
nanotechnologies. The results of this study may have far-reaching
consequence for nanoparticles applications.

\section{Acknowledgments}

We would like to thank Prof. H. K. Avetissian and Dr. G. F. Mkrtchian for
valuable discussions. This work was supported by State Committee of Science
of RA.


\section*{References}

\begin{thebibliography}{99}
\bibitem{1} Alivisatos A P 1996 \textit{J. Phys. Chem.} \textbf{100} 13226
--13239.

\bibitem{2} Zharov V P Kim Jin W Curiel D T Everts M 2005 \textit{%
Nanomedicine: Nanotechnology, Biology and Medicine} \textbf{1} 326-345.

\bibitem{3} Rotello V 2004 \textit{Nanoparticle: building blocks for
nanotechnology} (Springer) 300.

\bibitem{4} Neddersen J Gumanov G Cotton T 1993 \textit{Appl. Spectrosc.} 
\textbf{47} 1959.

\bibitem{5} Svrcek V Mariotti D Kondo M 2009 \textit{Opt. Express} \textbf{17%
}(2) 520--527.

\bibitem{6} Pyatenko A Shimokawa K Yamaguchi M Jishimura O Suzuki M 2004 
\textit{Appl. Phys. A} \textbf{79} 803-806.

\bibitem{7} Lalayan A 2005 \textit{Appl. Surface Science} \textbf{248}
209-212.

\bibitem{8} Lalayan A Avetisyan A Djotyan A 2005 \textit{Laser Phys. Lett.} 
\textbf{2}(1) 12--15.

\bibitem{9} Ganeev R A Baba M Ryasnyansky A I Suzuki M Kuroda H 2005 \textit{%
Appl. Phys. B }\textbf{80} 595--601.

\bibitem{10} Mafune F Kohno J Takeda Y Kondow T Sawabe H 2000 \textit{J.
Phys.Chem.} \textbf{B104} 9111.

\bibitem{11} Lalayan A 2014 \textit{Armenian Journal of Physics}\textbf{\ 7}%
(3) 122-126.

\bibitem{12} Men\'{e}ndez-Manj\'{o}n A Jakobi J Schwabe K Krauss J K
Barcikowski S 2009 \textit{J. of Laser Micro/Nanoeng.} \textbf{4} 95-99.

\bibitem{13} Bottcher C J F 1973 \textit{Theory of Electric Polarization }%
(Elsevier, Amsterdam) vol 1, 2nd ed.

\bibitem{14} Wolkin M V Jorne J Fauchet P M 1999 \textit{Phys. Rev. Lett.} 
\textbf{82}(1) 197-200.

\bibitem{15} Svrcek V Mariotti D Kondo M 2009 \textit{Opt. Express} \textbf{%
17}(2) 520-527.

\bibitem{16} Wang L Zunger A 1994 \textit{Phys. Rev. Lett.} \textbf{73}
1039. \ 

\bibitem{17} Wang L 2001 \textit{J. Phys. Chem. B }\textbf{105}(12)
2360-2364.
\end{thebibliography}
\end{document}